# Comet 2P/Encke in apparition of 2017: II. Polarization and color

Nikolai Kiselev [a,b,*], Vera Rosenbush [a,c], Oleksandra Ivanova [a,c,d], Ludmilla Kolokolova [e], Dmitry Petrov [b], Valeriy Kleshchonok [c], Viktor Afanasiev [f], Olena Shubina [a]

[a] *Main Astronomical Observatory of the National Academy of Sciences of Ukraine, 27 Zabolotnoho Str., 03143 Kyiv, Ukraine*
[b] *Crimean Astrophysical Observatory, Nauchnij, Crimea, Ukraine*
[c] *Astronomical Observatory of Taras Shevchenko National University of Kyiv, 3 Observatorna Str., 04053 Kyiv, Ukraine*
[d] *Astronomical Institute of the Slovak Academy of Sciences, SK-05960 Tatranská Lomnica, Slovak Republic*
[e] *University of Maryland, College Park, MD 20742, USA*
[f] *Special Astrophysical Observatory, Russian Academy of Sciences, 369167 Nizhnij Arkhyz, Russia*

We present results of imaging polarimetry of comet 2P/Encke performed on January 23, 2017 at the heliocentric (1.052 au) and geocentric (1.336 au) distances and phase angle 46.8°, 46 days before perihelion. Observations were made through the medium-band SED500 (λ5019/246 Å) and broadband r-sdss (λ6200/1200 Å) filters with the multimode focal reducer SCORPIO-2 at the 6-m BTA telescope of the Special Astrophysical Observatory (Russia). Dust in comet 2P/Encke was mainly concentrated in the near-nucleus region of the coma: the maximum dust/gas to leave $F_{em}/F_{cont}$ ratios were 1.5 and 2.9 in the SED500 and the r-sdss filters near the nucleus but dropped sharply to ~0.2 and ~1 at the distance ~2.500 km, respectively. Then these ratios began to increase at distances ~12,000 km from the nucleus, the ratio was ~0.3 (SED500) and ~1.3 (r-sdds). There were significant variations of polarization over the coma, which correlated with the variations in the dust color and dust/gas ratio. The maximum degree of polarization, ~8% in the r-sdss filter, was observed in the dust shell which was shifted by ~1.000 km towards the Sun. Polarization sharply dropped to ~4% at the distance ~3.000 km and then gradually increased with wave-like fluctuations with the distance from the nucleus, reaching ~8% at the distance ~12,000 km. A similar change in polarization was observed in the SED500 filter. After correction for gas contamination, using the dust/gas ratios from spectroscopy made on the same night, the values of polarization appeared to be ~4% in the near-nucleus region (~1.000 km), and reached 11–12% at the distance ~12,000 km in both filters. We also found an effect of nucleus polarization on the polarization of the dust coma in comet Encke in the r-sdss filter. The maximum value of the nucleus contamination was ~0.7%. Changes in polarization and color across the 2P/Encke coma indicate changes in physical properties of the dust particles with the dis-tance from the nucleus. Our Sh-matrix computer simulations of light scattering by Gaussian particles allow us to suggest that the observed trends in color and polarization are mainly result from changing particle size.

**Introduction**

Specific of the orbit of comet 2P/Encke (hereafter Encke) allows an intensive study its gas and dust coma, (see Newburn and Spinrad (1985), Gehrz et al. (1989), Sykes and Walker (1992), Fernández et al. (2005), Harmon and Nolan (2005), Sarugaku et al. (2015), Kwon et al. (2018), and our work Rosenbush et al. (2020) (hereafter referred as Paper I). Below we briefly review the results of the previous observations of comet Encke, which are necessary for discussion of the results obtained in this work.

Spectra of comet Encke show an extremely weak continuum in the visible region, and the dust/gas mass ratio was found to be 0.1 (Newburn and Spinrad, 1985). Therefore, comet Encke was classified as a gas-rich comet. In contrast, Reach et al. (2000) found that in the infrared range the dust/gas mass ratio was 10–30. A very weak overheating of the continuum and a weak silicate emission were interpreted by presence of the particles of radius larger than 5 to 10 μm in the cometary coma (Gehrz et al., 1989). On the base of the mid- and far-infrared data, Lisse et al. (2004) concluded that comet Encke emits a high fraction of big (20 μm and larger) grains, with the grain mass distribution similar to that for the interplanetary dust. There is a deficiency of micron-scale grains in comet Encke. A large number of millimeter-sized particles was found along the comet's orbit in the so-called dust trail (Reach et al., 2000).

Polarimetric observations of comet Encke were less extensive, although polarimetry greatly supplements photometric and spectral data by providing not only the scalar attribute of light (intensity or brightness), but also its "vectorial" parameters, i.e., the degree and orientation of polarization. These parameters are sensitive to such properties of dust particles as composition, size, shape, and structure, and their changes. Moreover, polarimetry can be used for a comet taxonomy based on properties of its dust (Chernova et al., 1993; Levasseur-Regourd et al., 1996; Kolokolova et al., 2007). As it is now generally accepted (see Kiselev et al., 2015 and references therein), there are two main classes of comets, which differ in their behavior near the maximum polarization, $P_{max}$: so-called high-$P_{max}$ and low-$P_{max}$ comets. This taxonomy, based mainly on the results of aperture (integral) polarimetry, is ambiguous. For example, Jewitt (2004) and Jockers et al. (2005) noted that the same comet could show both high and low degree of polarization depending on the measured coma area and spectral width of the filter. The polarization degree of the comet continuum can be affected by the comet activity, complex structure of the coma near the nucleus, and the molecular emissions, which depolarize the continuum, measured in the broadband filters. The situation became more complicated after Jewitt (2004) showed that the dust-poor comet Encke was highly polarized at phase angles 93–103° during the 2003 apparition. Jewitt concluded: "the reality of the two comet polarization classes remains unclear". At the same time, Jockers et al. (2005) demonstrated a significant difference in the polarization of comet Encke comparing the observations in the narrowband and broadband filters at phase angles 91–105°. The authors concluded that the gas contamination can significantly distort the results of the continuum polarization, especially for gas-rich comets.

Rather strong molecular emissions in the spectra of comet Encke create a good opportunity to study the polarization of various molecules. Using spectropolarimetric technique, Kwon et al. (2018) confirmed the effect of gas contamination and received the data on the polarization of the $C_2$ Swan, $NH_2$ α bands, and CN red system as well as for continuum at phase angle 75.7° of comet Encke in the 2017 apparition.

Comet Encke is the only dust-poor comet that can be regularly observed at large phase angles. In this paper, we present results of polarimetric observations of comet Encke carried out during observational campaign at the 6-m telescope of the SAO RAS in the 2017 apparition. Section 2 presents our polarimetric observations and data reduction, and Section 3 describes distribution of polarization parameters over the coma. In Section 4, we analyze observed polarization profiles over the coma, and then, in Section 5, define the effect of gas contamination on the continuum polarization. We discuss the variations of polarization and color over the coma in Section 6. The phase dependence of polarization for comet Encke is discussed in Section 7. The results of computer modeling of phase dependences of polarization are given in Section 8. Discussion of the obtained results and our conclusions are provided in Sections 9 and 10.

## Polarimetric observations and data reduction

Observations of comet Encke were carried out at the 6-m BTA telescope of the Special Astrophysical Observatory (Russia) on January 23, 2017 at the heliocentric (1.052 au) and geocentric (1.336 au) distances and phase angle 46.8°, 46 days before perihelion (March. 10.09, 2017). It was a single truly clear night for our observing period at the telescope. The seeing (FWHM) was 1.1″–1.2″ (1.114 km). The telescope tracked the motion of the comet to compensate its proper velocity during the exposures. A focal reducer SCORPIO-2 (Spectral Camera with Optical Reducer for Photometrical and Interferometrical Observations) installed at the primary focus of the 6-m BTA telescope was used in the long-slit spectroscopic, photometric and polarimetric modes (Afanasiev and Moiseev, 2011; Afanasiev and Amirkhanyan, 2012). Details of photometric and spectroscopic observations of comet Encke are described in Paper I. The back-illuminated CCD detector EEV 42–90 consisting of 2048 × 2048 px with a pixel size of 13.5 × 13.5 μm was used. The full field of view was 6.1′ × 6.1′ with a pixel scale of 0.18 arcsec/px, without binning.

Primary reduction of the data was performed using the IDL codes developed at the SAO RAS. The standard techniques of bias subtraction and flat-field correction were used. Binning 2 × 2 px was applied to the polarimetric images to improve the signal/noise (S/N) ratio of the measured signal. All polarimetric images were reduced to a single center using the isophotes method for combining the images. Note that the S/N ratio varied from about 140 in the near-nucleus coma to about 10 in the outer region of the coma, at distances 15,000–20,000 km.

The dichroic polarization analyzer (POLAROID) was used for the measurements of linear polarization of comet Encke in the broadband r-sdss (λ6200/1200 Å; hereafter, this way the central wavelength λ and FWHM of filters are given) and medium-band SED500 (λ5019/246 Å) filters. The analyzer was positioned in three fixed positions at the angles −60°, 0°, and +60°, what made up one measurement cycle. Using the intensities measured at those three positions of the POLAROID, $I(x,y)_0$, $I(x,y)_{-60}$ and $I(x,y)_{+60}$, we calculated the Stokes $q$ and $u$ parameters normalized to the total intensity $I$ in each point of the image for each cycle of exposures, according to the expressions:

$$I = \frac{2}{3}(I_0 + I_{+60} + I_{-60}),  \quad (1)$$

$$Q = \frac{2}{3}(2I_0 - I_{+60} - I_{-60}), \quad q = \frac{Q}{I}100\% = \frac{2I_0 - I_{+60} - I_{-60}}{I_0 + I_{+60} + I_{-60}}100\%, \quad (2)$$

$$U = \frac{2}{\sqrt{3}}(I_{+60} - I_{-60}), \quad u = \frac{U}{I}100\% = \sqrt{3}\,\frac{I_{+60} - I_{-60}}{I_0 + I_{+60} + I_{-60}}100\%. \quad (3)$$

The degree of polarization $P$ and the position angle of the polarization plane $\theta$ can be determined from the following expressions:

$$P = \sqrt{q^2 + u^2}, \quad (4)$$

$$tg2\theta = \frac{U}{Q} = \frac{u}{q} = \sqrt{3}\,\frac{I_{+60} - I_{-60}}{2I_0 - I_{+60} - I_{-60}}, \quad (5)$$

$$\theta = 0.5 arctg\left(\sqrt{3}\,\frac{I_{+60} - I_{-60}}{2I_0 - I_{+60} - I_{-60}}\right) + \Delta PA, \quad (6)$$

where $I_0$, $I_{+60}$, and $I_{-60}$ are reduced and corrected for sky background the intensity of the comet, and $\Delta PA$ is a correction for the zero point of instrumental position angle. A total measurement of the comet in one filter consisted of 5 cycles. Following (Clarke, 2010), we calculated the average values $\overline{P}$ and $\overline{\theta}$ using average values $\overline{q}$ and $\overline{u}$ computed for the entire series of observations:

$$\overline{P} = \sqrt{\overline{q}^2 + \overline{u}^2}, \quad (7)$$

$$\overline{\theta} = 0.5 arctg(\overline{u}/\overline{q}) + \Delta PA. \quad (8)$$

The uncertainties of $\overline{P}$ and $\overline{\theta}$ were estimated according to (Shakhovskoi, 1971) as:

$$\sigma_{\overline{P}} = \sqrt{0.5(\sigma_{\overline{q}}^2 + \sigma_{\overline{u}}^2)}, \quad (9)$$

$$\sigma_{\overline{\theta}} = 28.65° \, \sigma_{\overline{P}}/\overline{P}, \quad (10)$$

where $\sigma_{\overline{q}} = \sqrt{\sum(q_i - \overline{q})^2/(n-1)}$ and $\sigma_{\overline{u}} = \sqrt{\sum(u_i - \overline{u})^2/(n-1)}$ are values of the sample standard deviation of the parameters $\overline{q}$ and $\overline{u}$. Expressions (1–10) are applicable for each point of the image as well as for the average values of fluxes for specific areas of the coma.

The instrumental polarization, determined from observations of unpolarized standards (Serkowski, 1974), was <0.1% and was taken into account. The correction for the zero point of instrumental position angle $\Delta PA$ was found from observations of standard stars with large interstellar polarization (Hsu and Breger, 1982). The image processing, reduction, errors estimation, and the method of calculation of polarization parameters with SCORPIO-2 are described in more details by Afanasiev and Amirkhanyan (2012), Afanasiev et al. (2014), Ivanova et al. (2015, 2017a, 2017b), and Rosenbush et al. (2017).

The log of polarimetric observations of comet Encke are presented in Table 1, where we list the date of observation (the mid-cycle time, UT), the heliocentric ($r$) and geocentric ($\Delta$) distances, the phase angle (Sun-Comet-Earth angle) ($\alpha$), the position angle of the antisolar direction (scattering plane) ($\varphi$), the pixel size at the distance of the comet ($D$), the filter, the total exposure time during the night ($T_{exp}$), and number of cycles of exposures obtained in one night ($N$).

**Distribution of polarization parameters over the coma**

Fig. 1 shows the maps of linear polarization degree of comet Encke in the SED500 and r-sdss filters. The degree of polarization greatly varies (from ∼4% to ∼7%) over the coma which has a complex morphology in the polarized light. A fan-shaped structure is seen in the sunward hemisphere in both images, although it does not exactly coincide with the sunward direction. As it was shown in Paper I, this fan is composed of gas and dust. There is also a shell (or cloud of particles) with higher polarization (∼8%) at the distance about 1000 km from the nucleus. It is visible in both polarization images, but is particularly pronounced in the r-sdss filter. It seems that there is also a short jet-like feature ($PA \approx 290°$) with a higher polarization, extending from the near-nucleus region in the westward direction. This feature is also seen in Fig. 6 of Paper 1. The directions to the shell and jet-like feature (white line) inside the fan are shown only in the r-sdss images (Fig. 1) not to overload the figure.

The averaged degree $\overline{P}$ and position angle $\overline{\theta}$ of linear polarization of comet Encke in different apertures centered on the nucleus of the comet are given in Table 2 for the SED500 and r-sdss filters. It is well known (e. g., Kiselev et al., 2015) that the degree of polarization of light scattered by cometary dust generally increases with wavelength. Besides, in the red domain of the cometary spectrum, the contribution of depolarizing gas emissions is smaller. Therefore, as can be seen in Table 2 and Fig. 1, the degree of polarization in the r-sdss filter on average is higher than in the SED500 filter.

As Table 2 demonstrates, with increase of the aperture radius, the degree of polarization slightly decreases in both filters. A similar decrease of polarization degree with increasing coma area from 2.000 × 2.000 km to 10,000 × 10,000 km was also previously observed in comet Encke by Kiselev et al. (2004) in the red broadband filter (λ6940/790 Å) at phase angles 51.1° and 80.5°. The decrease in polarization with an increase of the coma area is likely due to a decrease of the dust/gas ratio at the distances ≤5000 km from the nucleus (see Section 4 for more detail).

The mean value of the position angle of polarization plane $\overline{\theta}$ is equal to 155.4° ± 0.3° which is almost orthogonal to the scattering plane: $\varphi = 59.6°$, that is $\Delta\phi = \overline{\theta} - \varphi = 95.8°$. Usually at phase angles larger than the inversion angle, $\alpha_{inv} \approx 22°$, $\Delta\phi \cong 90°$ and the cometary polarization plane is perpendicular to the scattering plane, thus, the polarization is positive.

Due to the effect of averaging polarization parameters in the coma, the aperture dependence of polarization parameters is not very informative, and it is more useful to consider polarization in specific regions of the coma. Fig. 2 presents distribution of polarization vectors over the coma. The measurements of polarization were performed within the 3 × 3 px (1.046 × 1.046 km) size area with the step 2 px (697 km). The degree of polarization and the position angle of polarization plane are represented by the length and orientation of the bars. The observed distribution of polarization vectors is somewhat different from that which was observed during apparition in 2003 (Kiselev et al., 2004).

Fig. 2 shows that there are systematic differences in polarization degree between different areas along the fan ($PA \approx 290°$) in both filters. The degree of polarization decreases from the optocenter, reaching a minimum (∼4%) at the distance of about 2000 km, and then increases to ∼8% at ∼8000 km from the nucleus. A similar change in polarization is seen in Fig. 1 (see also Fig. 3, Section 3). On average, the position angle of polarization plane is 155.2° ± 1.8° and 155.4° ± 1.6° in the SED500 and r-sdss filters, respectively. Thus, the systematic difference between the position angles of polarization plane and the scattering plane is about 96°, and the polarization of comet Encke over the whole coma is positive. The uncertainty of measurement of polarization plane in each area at the level of 1 $\sigma$, i.e., is about 3. However, in some coma regions, the polarization plane deviation from average may exceed 3°σ. In previous works, Mirzoyan and Khachikian (1959), Martel (1960), and Clarke (1971) studied a distribution of polarization vectors in the comae and tails of comets. These authors revealed both insignificant systematic deviations of polarization planes from the scattering plane and local regions of coma with significant differences 0° ≤ $\Delta\phi$ ≤ 90°. Clarke explained this effect by light scattering on aligned elongated particles. Gnedin et al. (1999) discussed how this mechanism can affect the plane of polarization depending on the dust particle properties, specifically, their size. Thus, inhomogeneity of the dust distribution in the coma should produce inhomogeneity in the position of polarization plane.

**The observed polarization profiles**

Fig. 3 displays the radial profiles of polarization across the images obtained in the SED500 and r-sdss filters. The cuts were made in sunward (Sun) and perpendicular to the Sun directions (Sun_p), as well as along the fan (fan) and perpendicular to the fan directions (fan_p). The measurements of the polarization were performed with 3 × 3 px size aperture (or 1046 × 1046 km at the comet) with the 2 px step (697 km) along the selected directions in the coma with increasing distance from the nucleus. Fig. 3 also shows the expected degree of polarization (3.95%) for molecular emission $C_2$ at phase angle 46.8° (horizontal dashed lines), according to Öhman (1941) and Le Borgne et al. (1987a, 1987b). The uncertainties in polarization degree vary from ∼0.2% near the optocenter to ∼0.7% at coma periphery that is consistent with a decrease of the signal-to-noise ratio by about 14 times.

As one can see, within the limits of measurement errors, there are no significant differences, in radial profiles of polarization in different directions on the solar side. This means that the dust particles experienced the same changes in all directions. At the same time, significant changes occurred in the radial distribution of polarization in the coma: the polarization in the near-nucleus area (∼1000 km) was significantly higher than that in the surrounding coma. It dropped sharply from ∼6%

**Table 1**
Log of the polarimetric observations of comet 2P/Encke.

| Date, UT 2017 January 23 | $r$ (au) | $\Delta$ (au) | $\alpha$ (deg) | $\varphi$ (deg) | $D^a$ (km/px) | Filter | $T_{exp}$ (sec) | N |
|---|---|---|---|---|---|---|---|---|
| 15:43:45–16:04:58 | 1.052 | 1.336 | 46.8 | 59.6 | 348.8 | r-sdss | 900 | 5 |
| 16:07:17–16:28:32 | 1.052 | 1.336 | 46.8 | 59.6 | 348.8 | SED500 | 900 | 5 |

[a] Taking into account the binning 2 × 2 px.

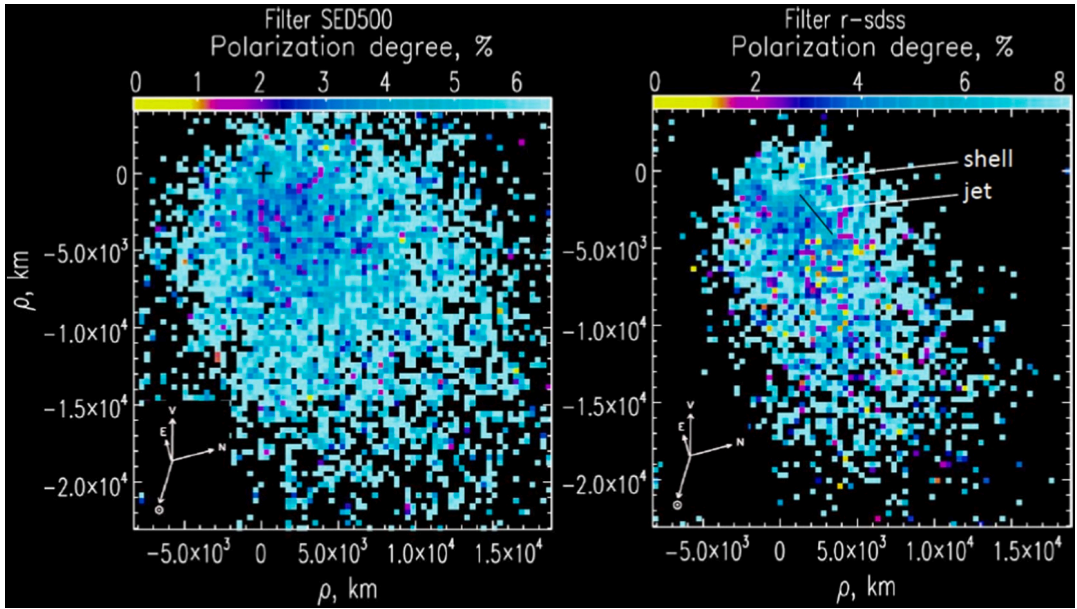

**Fig. 1.** Distribution of linear polarization degree over the 2P/Encke coma on January 23, 2017. Polarization maps are for the SED500 (left-hand) and r-sdss (right-hand) filters. The locations of the shell and jet-like structure (white line) are shown in the r-sdss image on the right. The optocenter is marked with a white cross. North, East, sunward, and velocity vector directions are indicated. Negative distance is in the solar direction.

**Table 2**
Parameters of linear polarization of comet 2P/Encke measured in different apertures and filters.

| Aperture radius[a] (km) | Filter SED500 | | Filter r-sdss | |
|---|---|---|---|---|
| | $\overline{P} \pm \sigma_{\overline{P}}$ (%) | $\overline{\theta} \pm \sigma_{\overline{\theta}}$ (deg) | $\overline{P} \pm \sigma_{\overline{P}}$ (%) | $\overline{\theta} \pm \sigma_{\overline{\theta}}$ (deg) |
| 704 | 5.7 ± 0.3 | 155.9 ± 1.3 | 6.2 ± 0.3 | 154.6 ± 1.2 |
| 1759 | 5.2 ± 0.3 | 156.1 ± 1.6 | 6.5 ± 0.3 | 154.9 ± 1.3 |
| 3519 | 5.1 ± 0.4 | 155.1 ± 2.0 | 5.9 ± 0.4 | 156.1 ± 1.8 |
| 4223 | 5.0 ± 0.4 | 154.1 ± 2.2 | 5.8 ± 0.4 | 156.2 ± 2.0 |

[a] HMHW of seeing is 557 km.

(SED500 filter) and from 7.5% (r-sdss filter) down to ~4% in both filters at ~3000 km distance from the nucleus, and then increased, with wave-like variations, up to ~7% towards the edge of the coma. Note also that the maximum degree of polarization of comet Encke in the r-sdss filter is shifted relatively to the photometric center of the coma along the fan to the distance ~1000 km and some increase of polarization is observed at distances about 2000 km in the anti-solar direction. Increasing polarization in this area (shell?) may manifest newly released dust grains, which have different physical properties than older material still present within the field of view. In addition, it is likely that the dust was affected by the prominent fan observed on the sunward side and, hence, was not symmetrically distributed in the near-nucleus area of the coma. To explain the observed variations of the polarization degree over the coma, it is necessary to obtain the polarization in the continuum, i.e. to remove less polarized (see Fig. 3) molecular radiation penetrated in the filters, causing so-called "gas depolarization" or "gas contamination".

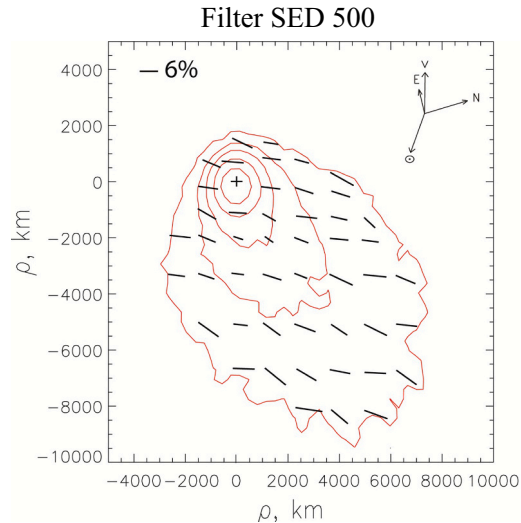

**Fig. 2.** Distribution of the polarization vectors in the coma of comet 2P/Encke on January 23, 2017 in the SED-500 filter. The isophotes are superimposed on the images. The orientation of the vectors indicates the direction of the local polarization plane, and the length of the vectors indicates the degree of polarization. The arrows point in the direction to the Sun (☉), North (N), East (E), and velocity vector of the comet as seen in the observer's plane of sky (V).

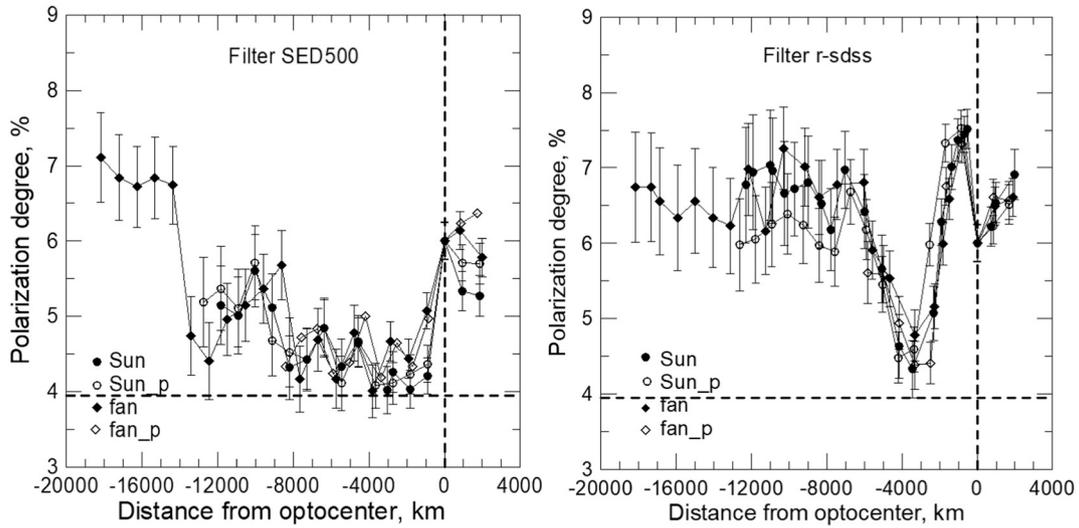

**Fig. 3.** The observed radial profiles of polarization across the coma of comet 2P/Encke in different directions as a function of the distance from the nucleus which is indicated by vertical dashed line in the SED500 (left-hand) and r-sdss (right-hand) filters. The horizontal dashed lines are the expected degree of polarization for the two-atomic molecular emissions equal to 3.95% at $\alpha = 46.8°$ (according to Öhman, 1941; Le Borgne et al., 1987a, 1987b). Negative distance is in the solar direction.

### Contamination of dust coma polarization by gas emissions and nucleus

Due to faintness of comet Encke, we could not obtain its polarization images in the narrowband continuum filters. However, we can remove the influence of the gas contamination in the medium-band SED500 ($\lambda 5019/246$ Å) and broadband r-sdss ($\lambda 6200/1200$ Å) filters using the spectral observations obtained at the same observational night and described in Paper I. For this, we measured the continuum and emission fluxes along the spectrograph slit projected on the coma. According to the spectrum of comet Encke, the SED500 filter ($\lambda 5019/246$ Å) transmits the radiation scattered by dust and strong $C_2(0,0)$ ($\lambda 5160$ Å) emission, whereas the r-sdss filter ($\lambda 6200/1200$ Å) transmits the radiation from dust and less strong $NH_2(0,7,0)$ ($\lambda 5750$–$5810$ Å) and $C_2(0,1)$ ($\lambda 5635$ Å) emissions. Therefore, a significantly greater effect of gas contamination on the polarization is in the filter SED500. This can be seen in the polarization images in Figs. 1 and 3 and Table 2.

To evaluate the gas contamination effect, we used the ratio of continuum to $C_2$ ($\lambda 5140$ Å) emission in the SED500 filter as well as the ratio of continuum to $C_2$ ($\lambda 5635$ Å) + $NH_2$ ($\lambda 5750$–$5810$ Å) emission in the r-sdss filter along the spectrograph slit (see Paper I). Here, the continuum means the total radiation scattered by the dust particles of the coma and cometary nucleus, that is $F_{cont}(\rho) = F_{dust\ coma}(\rho) + F_{nucl}(\rho)$. In Fig. 4, we reproduce the ratio of fluxes $F_{cont}(\rho)/F_{em}(\rho)$ in the cuts along the spec-trograph slit in both filters at the distances up to $\sim$15,000 km. It is likely that the dust is mainly concentrated in the near-nucleus region of comet Encke where the dust/gas ratio $> 1$ for both filters. There is a sharp drop of the dust/gas ratio at the distances $< 1500$ km. The observed weakness of the emissions $NH_2$ and $C_2$ in the red domain of the spectrum indicates that the dust/gas ratio in the r-sdss filter remains greater than unity at larger distances from the nucleus. The strong emission $C_2$, penetrating into the SED500 filter, remains predominant at considerable distances from the nucleus. As a result, the $F_{cont}/F_{em}$ ratio in this filter is $<1$. The continuum radiation that penetrates in this filter constitutes $\sim$0.23 of the total radiation. The presence of dust at large distances from the nucleus is unequivocally confirmed by the measured degree of polari-zation, which is higher than that of molecular emissions $P = 3.95\%$ (see Fig. 3). In general, an expression for the observed polarization in the filter transmitting emission and continuum, $P_{obs}(\rho)$, with the distance from the

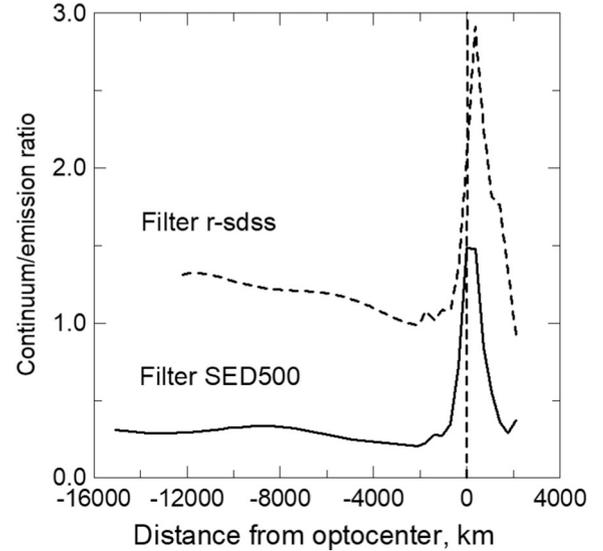

**Fig. 4.** The $F_{cont}/F_{em}$ flux ratios in the SED500 and r-sdss bands along the spectrograph slit ($PA = 74.6°$) projected on the coma of comet 2P/Encke as a function of the distance $\rho$ from the optocenter.

optocenter $\rho$ can be written as:

$$P_{obs}(\rho)(F_{cont}(\rho) + F_{em}(\rho)) = P_{cont}(\rho)F_{cont}(\rho) + P_{em}(\rho)F_{em}(\rho), \quad (11)$$

where $F_{cont}(\rho)$ and $F_{em}(\rho)$ are the fluxes of continuum and emission along the slit, $P_{cont}(\rho)$ is unknown degree of the pure continuum polarization. We assume that the degree of polarization in emission does not depend on the distance and can be found for the observed phase angle $\alpha$ from the Öman's expression (Öhman, 1941):

$$P(\alpha) = 100\% P_{90} sin^2(\alpha) / (1 + P_{90} cos^2(\alpha)), \quad (12)$$

where $P_{90} = 0.077$ is the maximum polarization for the $C_2$ emission at the phase angle $\alpha = 90°$ (Le Borgne et al., 1987a, 1987b). We assume that the degree of polarization in the $NH_2$ emission is close to that of diatomic molecule $C_2$, i.e. $P_{90} = 7.7\%$ (Jockers et al., 2005), although Kwon et al. (2018) found that the $NH_2(0,7,0)$ molecules have lower degree of polarization, $P_{90} = (4.0 \pm 1.0)\%$. At the $3\sigma$ level, this value is close to the result by Jockers et al. In relative units, $F_{cont}(\rho) + F_{em}(\rho) = 1$. Taking the ratios $k_1(\rho) = F_{cont}(\rho)/F_{em}(\rho)$ from Fig. 4 and Eq. (11), we obtain the continuum polarization profiles through the cometary coma along the slit using the following equation:

$$P_{cont}(\rho) = \frac{P_{obs}(\rho)(k_1(\rho) + 1) - P_{em}}{k_1(\rho)}, \qquad (13)$$

where $P_{em} = 3.95\%$ at the phase angle $\alpha = 46.8°$. One can see that the polarization of pure continuum in both filters is systematically higher than the observed polarization degree, excluding the coma regions with minimal polarization: $P_{min} = 3.9 \pm 0.4\%$ at the distance $\rho \approx 1.500$ km for the SED500 filter and $P_{min} = 4.4 \pm 0.4\%$ at $\rho \approx 3.000$ km for the r-sdss filter.

The removal of the gas contamination will inevitably introduce additional errors in polarization degree. Using the absolute error of the function of several approximate arguments (Schigolev, 1969), we can derive the average additional error:

$$\sigma_{Pcont} \leq \Delta P_{cont} = \sum |\frac{\partial f}{\partial x_i}|\Delta x_i, \qquad (14)$$

where $\Delta P_{cont}$ is absolute error of polarization degree, $\frac{\partial f}{\partial x_i}$ is partial derivatives of function (13), $\Delta x_i$ represents absolute errors of arguments $P_{obs}$ and $k_1$. For simplicity, we do not consider any dependence of the used quantities on the cometocentric distance. Taking into account that measurement error of the $F_{cont}$ and $F_{em}$ fluxes is 10% and $k_1 \approx 1$, we obtain the error of $\Delta k_1 = 0.2$. In the cuts, the average values of $P_{obs}$ and its error $\Delta P_{obs}$ are, respectively, 5% and 0.4% for the SED500 filter and 6% and 0.4% for the r-sdss filter. Therefore, the absolute error $\Delta P_{cont}$ will be 1% and 1.2% for the SED500 and r-sdss filters, respectively. Absolute errors limit the maximum values of uncertainties. These additional systematic uncertainties are shown by bars for each filter in Fig. 5.

The corrected polarization profiles show some features in the spatial structure of polarization in the coma. The maximum degree of polarization is slightly offset relative to the comet's optocenter towards the Sun in both filters that may be explained by presence of dust shell (or dust cloud) near the nucleus in which there were some highly polarizing dust particles. The maximum polarization in the near-nucleus area increases from ~6% up to ~8% and from ~7.5% up to ~11% in SED500 and r-sdss filters, respectively. Then it drops sharply to ~4% at the distances ~1.500 km and ~3.000 km, and after that gradually increases with the distance, reaching ~12% and ~11% at 12,000–15,000 km. A similar effect was observed in the dust-rich comet 67P/Curyumov-Gerasimenko (thereafter 67P/C-G) at the phase angle about 32°: polarization of this comet sharply decreased within the first 5.000 km from ~8% to ~2% and then gradually increased reaching ~7% at 36,000 km (Rosenbush et al., 2017).

Using the results of Paper I, we can also take into account the contribution of the polarization of light scattered by the cometary nucleus in the polarization of the dust coma. In Fig. 6, we reproduce the ratio of fluxes $k_2(\rho) = F_{nucl}(\rho)/F_{dust\,coma}(\rho)$ in the cut along the spectrograph slit in the r-sdss filter. It is seen in the figure, that the maximum of the $F_{nucl}/F_{dust\,coma}$ flux ratio is shifted about 700 km away from the Sun.

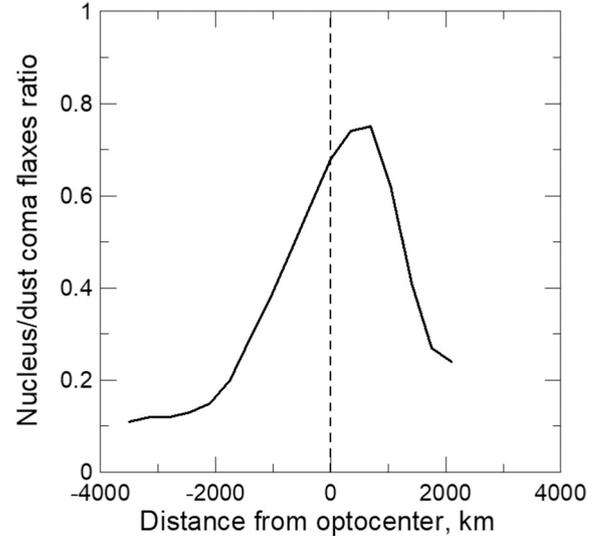

**Fig. 6.** The $F_{nucl}/F_{dust\,coma}$ flux ratio in the r-sdss band along the spectrograph slit ($PA = 74.6°$) projected on the 2P/Encke coma as a function of the distance $\rho$ from the optocenter.

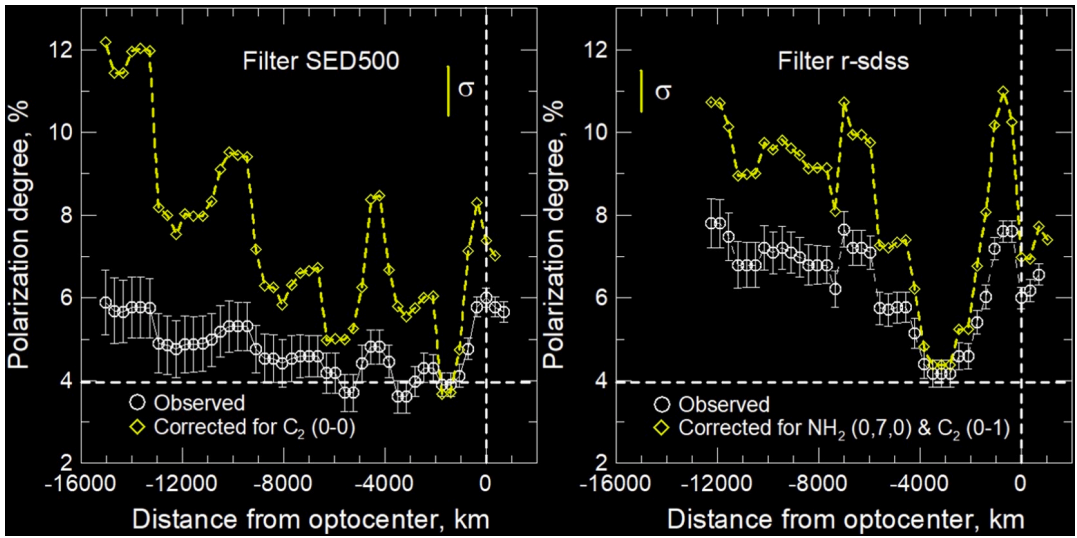

**Fig. 5.** The observed and corrected for gas contamination radial profiles of polarization along the spectrograph slit located at the position angle of 74.6° as a function of the distance from the optocenter in the SED500 (left-hand) and r-sdss (right-hand) filters. The uncertainty in observed polarization in the optocenter is ~0.2%, while it is ~0.7% at coma periphery for both filters. The systematic uncertainties in the corrected polarization (shown by blue lines and diamonds) are ≤1% and ≤ 1.2% for the SED500 and r-sdss filters; they are shown by the vertical lines. The horizontal dashed lines are the degree of polarization for the diatomic molecules equal to 3.95% at $\alpha = 46.8°$, according to Eq. (12) and Le Borgne et al. (1987a, 1987b). (For interpretation of the references to color in this figure legend, the reader is referred to the web version of this article.)

To find out the polarization degree of the nucleus at the observed phase angle, we used the measurements of the polarization degree of the nucleus of comet Encke in the R filter by Boehnhardt et al. (2008) and Jewitt (2004). Using these data, the polarization phase curve was obtained and fitted with the trigonometric expression suggested by Lumme and Muinonen (1993):

$$P(\alpha) = b(sin\alpha)^{c_1}(cos(\alpha/2))^{c_2}sin(\alpha - \alpha_0), \quad (15)$$

where $b$, $c_1$, $c_2$, and $\alpha_0$ are fitting parameters. From Eq. (15), we found the polarization degree of the nucleus of comet Encke equal to 6.35% for the phase angle $\alpha = 46.8°$ Computation of the polarization contribution of light scattered by the nucleus in the observed polarization of the coma was performed according to the formula:

$$P_{dust\ coma}(\rho) = P_{obs}(\rho)(1 + k_2(\rho)) - 6.35k_2(\rho) \quad (16)$$

The same Eq. (14) can be used to estimate the error in polarization degree introduced by the procedure for taking into account nucleus brightness according to Eq. (16). Assuming for the optocenter $P_{obs} = 6\%$, $\Delta P_{obs} = 0.2\%$, $k_2 = 0.7$, and $\Delta k_2 = 0.1$, we obtain the error in polarization curve corrected for nucleus brightness, $\sigma_{Pdust\ coma} \leq \Delta P_{dust\ coma} = 0.37\%$.

The final expression for the polarization of the dust coma that accounts for the total flux contribution of the main molecular emissions and the nucleus is the following:

$$P_{dust\ coma}(\rho) = (P_{obs}(\rho) - P_{em}(\rho)F_{em}(\rho) - P_{nucl}(\rho)F_{nucl}(\rho))/F_{dust\ coma}(\rho), \quad (17)$$

where $F_{dust\ coma}(\rho) + F_{nucl}(\rho) + F_{em}(\rho) = 1$. Fig. 7 shows the variations of polarization along the spectrograph slit projected on the coma in the r-sdss filter: the observational data are indicated by black circles; the data after removing the gas contamination are blue diamonds, the nucleus contamination is shown by pink squares, and a combined contamination by the nucleus and gas is denoted by red circles. Within the innermost region, the contribution to the polarization of the coma from molecular emissions was ~4% relative to the observed polarization, and ~0.7% from the nucleus; their total action was ~5%. At the distances larger than ~1000 km from the optocenter of coma, the contribution of the scattered light from the nucleus in the coma polarization was negligible.

## 6. Variations of polarization and color of dust over the coma

The radial variations of polarization and BC–RC color in comet Encke with increasing distance from the nucleus are shown in Fig. 8 for January 23, 2017. The polarization was corrected for the emission and nucleus contribution. The BC–RC color, that is notated as BC'-RC' in Paper 1, corresponds to the color free of the gas and nucleus contaminations taken from Paper 1 (Fig. 11 there). The measurements of the polarization and color were performed in the 1.046 × 1.046 km area at the comet along the direction of the spectrograph slit on the corresponding images. It is seen that at the distances < 4.000 km the dust color very sharply decreases with the distance from the nucleus, from about $1.4^m$ to $0.4^m$. Variations of polarization with the distance from the nucleus are more complex, and these changes are similar in both filters (see Fig. 5). As in comet 67P/C-G, there is a "turning point" (see Fig. 8), where polarization changes its trend from a sharp decrease to gradual increase: it decreases within the first 3.000 km from ~13% to ~4% (r-sdss filter), and then gradually, although with some wave-like variations, increases, reaching ~11% at 12,000 km. As in the case of comet 67P/C-G (Rosenbush et al., 2017), there is some correlation between the variations of polarization degree and color index with the cometocentric distance. Note that Jewitt (2004) observed a similar behavior of polarization and color in the coma of comet Encke during its perihelion passage in 2003. The significant radial variations of the polarization, corrected for gas and nucleus contamination, and color of comet Encke may indicate only changes in physical properties of the dust particles.

## 7. Phase-angle dependence of polarization of comet Encke

The observed and corrected degree of polarization in the 1.046 × 1.046 km area of the coma at distance ~ 700 km from the optocenter along cuts as well as data obtained in different areas of the coma by Kiselev et al. (2004), Jewitt (2004), Jockers et al. (2005), and Kwon et al. (2018) (all data were mainly obtained in the red spectral range) were used to construct the phase-angle dependence of polarization for comet Encke (Fig. 9). The curves 1 and 2, which are average dependences of polarization for high-$P_{max}$ and low-$P_{max}$ comets, are taken from Kiselev et al. (2015). Curve 3 is the phase-angle dependence of polarization for two-atomic molecules, according to Eq. (12) and (Le Borgne et al., 1987a, 1987b). In the broadband filters, the degree of polarization measured in a large coma area (large symbols) is close to the values of polarization for two-atomic molecules (curve 3).

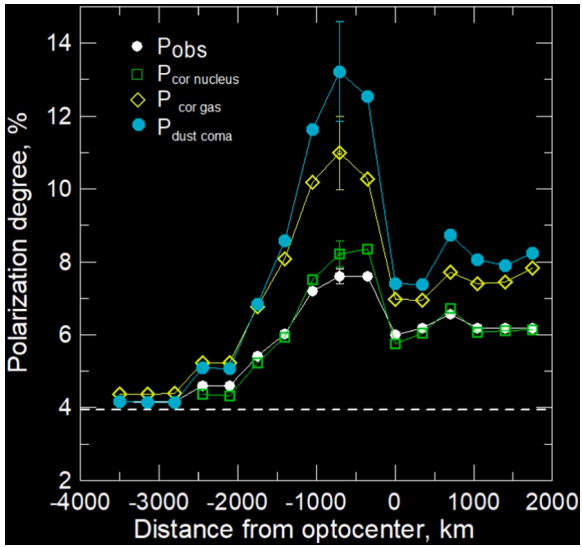

**Fig. 7.** Radial profiles of the 2P/Encke polarization along the spectrograph slit located at the position angle of 74.6° as a function of the distance from the optocenter in the r-sdss filter: the observed profiles (black circles), corrected for the nucleus contribution (pink squares), gas contamination (blue diamonds), and total gas + nucleus (red circles). The error bars are shown for the maximum value of polarization. The horizontal dashed line is the expected degree of polarization for the diatomic molecules equal to 3.95% at $\alpha = 46.8°$. The inherent polarization of light scattered by the nucleus of comet Encke is 6.35% in the r-sdss filter. (For interpretation of the references to color in this figure legend, the reader is referred to the web version of this article.)

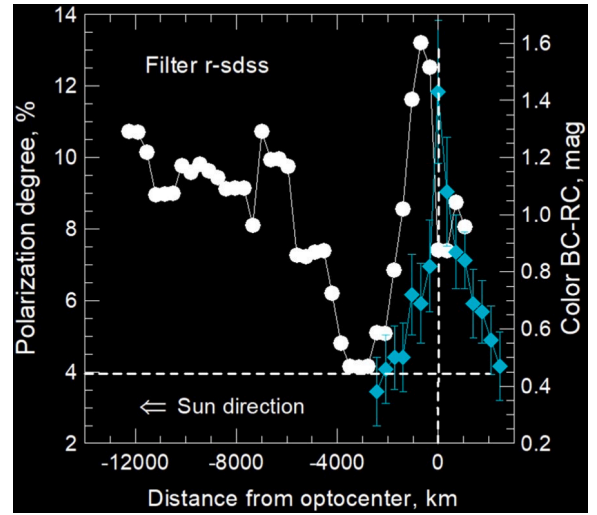

**Fig. 8.** The profiles of polarization (black line with circles) and BC–RC color (red line with diamonds) corrected for the emission and nucleus contributions over the coma of comet 2P/Encke in the direction of the spectrograph slit. For simplicity, only polarization in the r-sdss filter is shown (left axis). The uncertainty in the corrected polarization curve is the same as in Fig. 5. The right axis is color BC–RC taken from Paper I (Fig. 11 there). (For interpretation of the references to color in this figure legend, the reader is referred to the web version of this article.)

Our measurements of polarization of comet Encke (open triangles) are located between curve 3 and the low-$P_{max}$ comets (curve 2). After correcting for the gas contamination, the polarization data (solid triangles) became closer to the curve for the high-$P_{max}$ comets (curve 1). The resulting difference in polarization is not significant (only 2.4% and 3.4% in the SED500 and r-sdss filters), however, it exceeds the uncertainty of the corrected data, which is about 1% (see Fig. 5). The data by Kwon et al. (2018), obtained only a month after our observations, showed a high degree of polarization of comet Encke's dust at phase angle 75.7°, comparable to the results of previous observations by Jewitt (2004) and Jockers et al. (2005). Thus, Fig. 9 confirms that the inclusion of comet Encke in the class of low- or highly-polarized comets depends on the contribution of gas emissions to the measured continuum, which in turn is determined by the width of the used filters and the coma region.

## 8. Sh-matrix modeling of phase-angle dependence of polarization

Above, we showed that the depolarizing effect of gas, depending on the dust/gas ratio, is a real effect. However, gas contamination is not the main reason of the change of the polarization with distance from the nucleus. If the correction for gas contamination was performed correctly, then, within the limits of measurement errors, the behavior of the corrected polarization and color depends only on the changing properties of the dust.

Up to about 3.000 km, the polarization drops sharply from ~11%

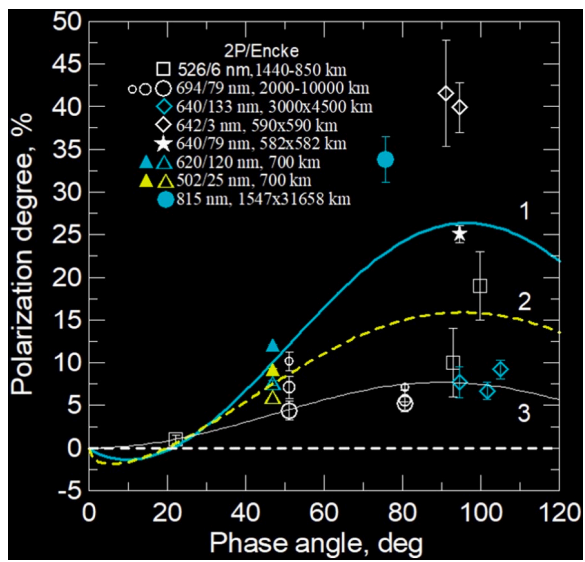

**Fig. 9.** Polarization of comet 2P/Encke versus phase angle. Aperture data were obtained by Kiselev et al. (2004) (circles), Jewitt (2004) (squares), Jockers et al. (2005) (diamonds, asterisk), Kwon et al. (2018) (filled circle), and this work (open and filled triangles for observed and corrected data, respectively). The composite polarization curves for the high-$P_{max}$ and low-$P_{max}$ comets taken from Kiselev et al. (2015) are denoted as 1 and 2, respectively. Curve 3 is the phase-angle dependence of polarization for two-atomic molecules, according to Eq. (12) and (Le Borgne et al., 1987a, 1987b).

(after taking into account the nucleus contribution) to ~4.5% in the r-sdss filter and from ~8% to ~4.2% in the SED500 filter. At the same time, color of the dust became bluer. The reasons may be a rapid change in the size and/or composition as well as the structure of the dust particles due to the loss of volatiles. At distances larger than ~3.000 km, the degree of polarization increases from ~4.2% to 11–12%. Again, if the dust/gas ratio and gas contamination are taken into account correctly, these changes in polarization are due to changes in the properties of dust. In order to identify what properties of dust particles can lead to the observed effects, we conducted computer simulations.

Rosenbush et al. (2017) modeled change in color and polarization in comet 67P/C-G with changing particle size and could qualitatively simulate the observed trends for a size distribution of spherical particles. Recently, Kolokolova and Petrov (2017) presented results of the modeling of those trends using Sh-matrix method. Being based on the T-matrix technique, this approach separates the shape-dependent factors from the size- and refractive index dependent factors, presenting the shape with a shape matrix, or Sh-matrix (Petrov et al., 2011, 2012). Sh-matrix method keeps all advantages of the T-matrix method (Mishchenko et al., 1996), including analytical averaging over particle orientation. Additionally, in this method size and refractive index dependences are incorporated through analytical operations on the Sh-matrix to produce the T-matrix elements. The Sh-matrix elements themselves are rather simple functions and can be solved analytically for any shape of particles. This makes Sh-matrix approach an effective technique to simulate light scattering by particles of complex shape and surface structure.

Specifically, the Sh-matrix approach was found to be very convenient to consider light scattering by Gaussian particles, i.e. the particles created by disturbing a sphere of a given radius the way that the sphere radii become randomly lognormally distributed (Muinonen, 1996). Two radial distances relate to one another through "correlation angle", changing which from 0 to 90° we can model a variety of particles from spheres to particles of a random complex shape. We consider cometary dust as Gaussian particles with the correlation angle 20°. Fig. 8 shows the shape of the considered Gaussian particles and the results of our Sh-matrix computations. We build the particles of amorphous silicates (the refractive index $m = 1.689 + 0.0031i$ at $\lambda 4500$ Å and $m = 1.677 + 0.0044i$ at $\lambda 6500$ Å (Scott and Duley, 1996)) and Halley-like dust, i.e., the dust made of silicates, amorphous carbon, and organics, the refractive index $m(\lambda 4500$ Å$) = 1.88 + 0.47i$, $m(\lambda 6500$ Å$) = 1.98 + 0.48i$ (Kimura et al., 2003). The Halley-like particles are considered to be porous, with the porosity varying from 70% to 90%; their refractive index was calculated using the Maxwell Garnett rule.

Fig. 10 shows that change in the size of particles can explain the observed trends for a variety of porosities and refractive indexes; the best fit was obtained for a mixture consisting of 90% of porous Halley-like particles and 10% of silicate particles. From the left panel of the figure, which shows the maps of polarization as a function of phase angle (horizontal axis) and particle radius (vertical axis), one can see that for the phase angle of our observation as the size of the scattering particles decreases, the polarization decreases. However, at the radius ~ 0.5 μm, as the scattering by the particles approaches the Rayleigh regime, the trend reverses and polarization starts increasing. The right panel represents a similar map for the color of particles. It demonstrates that a decrease in the size of the scattering particles causes a gradual change of color from red to blue. For both, polarization and color, the modeled trends are consistent with the observational data shown in Fig. 8.

Note that we also performed the simulations varying composition of the particles to model decreasing content of organics as particles move out of the nucleus. This could reproduce the color changes, but we could not reproduce an increase in polarization if we keep the size of particles unchanged. Change in particle porosity, which could increase with the distance from the nucleus, does not produce any regular change in color. Note that our goal was to check how realistic are the observed trends, not to fit them quantitatively as our particles are not as complex as aggregated particles that represent the dominating component of the cometary dust.

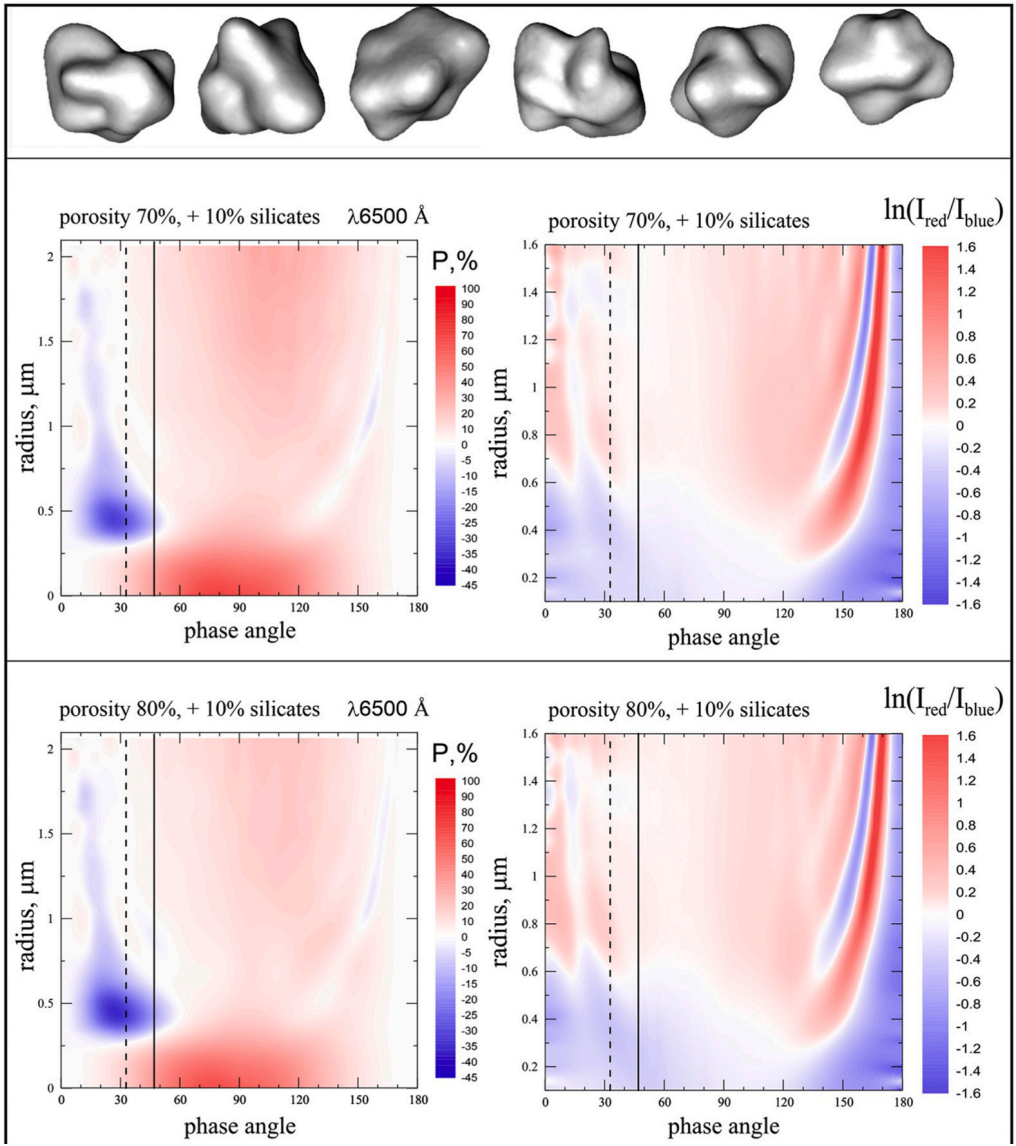

**Fig. 10.** Results of the Sh-matrix modeling of phase dependences of polarization (left panel) and color (right panel). Vertical axis shows radius of particles. All plots are for a mixture of 90% of porous Halley-like particles and 10% of solid silicate particles. Halley-like particles have porosity 70% (top panel) and 80% (bottom panel). Solid black line shows the phase angle of our 2P/Encke observations, dashed line is for the observations of 67P/Churyumov-Gerasimenko described in Rosenbush et al. (2017). On the top, we show the Gaussian particles used for the modeling.

## Discussion

Comet 2P/Encke is an ideal comet for studying the effect of gas contamination on the continuum polarization because it is practically the only gas-rich comet that can be observed at large phase angles where this effect strongly manifests itself. A puzzled picture is produced by its polarization data, according to which Encke can be referred to both, high-$P_{max}$ and low-$P_{max}$ classes of comets.

*Dust coma polarization, nucleus contribution, and gas contamination*

This paper is the first attempt to study the effect of the nucleus polarization on the coma polarization. For comet Encke, this effect is small, ≤0.7% at phase angle of 46.8°, and is observed directly near the cometary nucleus, at distances ≤1.000 km. As it was shown in Section 5, after correcting for the nucleus and gas contamination, the degree of polarization at phase angle of 46.8° in the near-nucleus region increased by ~2%–5% depending on the filter. The gas contamination effect is much larger at large phase angles (see Fig. 9). Jockers et al. (2005) found that the observed polarization of comet Encke in the λ6621/59 Å filter at $\alpha = 94.6°$ changed from 22.7% to 31.1% after correcting for the NH$_2$ emission. The reliability of these data was confirmed by their consistency with the high polarization degree, $P \approx 35\%$, measured in the continuum (λ5259/52 Å) at $\alpha = 96°$ by Jewitt (2004). Also, only a month after our observations, a high polarization degree of this comet, $P = 33.8\%$ at $\alpha = 75.7°$, was obtained by Kwon et al. (2018) after correcting the continuum data (λ8150 Å) for gas contamination. Obviously, the degree of polarization of comet Encke in the near-nucleus coma region at large phase angles is 30–35%, which is considerably higher than the maximum polarization of most dusty comets (see Kiselev et al., 2015 and references therein). The depolarizing effect of molecular emissions results from a much smaller maximum of polarization degree for them (7.7%) caused by resonance fluorescence in comparison with the continuum polarization produced by the scattered solar light on dust particles (~30%). The depolarizing effect of gas contamination can be noticeable even for gas-poor comets if the polarization was measured in the broadband filters U, B, V, and even in the broadband red filters, transmitting some fraction of the slightly polarized molecular emissions of CN, C$_2$, and NH$_2$. This effect becomes negligible at phase angles smaller than 30° where the phase dependence of polarization of molecular emissions approaches zero (see Fig. 9).

*Diversity in comet polarization*

As we mentioned above, there are two groups of comets with different values of polarization maximum: the high-polarization comets (high-$P_{max}$) with polarization peak of about 28% at phase angle $\alpha \approx 95°$ and the low-polarization comets (low-$P_{max}$) with polarization peak of about 10–15% at $\alpha \approx 90°$. Understanding the reasons of this diversity may shed light on differences in the formation and/or evolution of these two types of comets.

Kiselev (1981), Dobrovolsky et al. (1986), Chernova et al. (1993), and Jockers (1997) considered the gas contamination as one of the main reasons of diversity in the polarization of comets. This conclusion was based on aperture observations of comets with filters, which simultaneously transmit radiation from the continuum and less polarized gas emissions. In this case, the polarization depends on the dust/gas ratio. However, the polarization depends also on the width of filter, strength of cometary continuum and emissions and their variability over time. As a result, in the "phase angle – polarization" diagram, comets with a strong continuum and relatively weak emissions will form the class of high-polarization comets, polarization of which weakly depends on the filter and the size of measured coma. Comets with a moderate/weak continuum and relatively strong emissions will form the class of low-polarization comets. In the "phase angle – polarization" diagram, the maximum polarization of such comets lies within the range from ~8% to ~20%; it can vary even for one comet, and strongly depends on the size of measured coma.

Dollfus and Suchail (1987), Hadamcik and Levasseur-Regourd (2009), and Shestopalov and Golubeva (2017) also recognized a reduction of the observed polarization caused by gas emissions. Important point is that the polarization depends on the dust/gas ratio, thus, even the comets with a significant amount of dust can show low $P_{max}$, if they also have a significant amount of gas.

Chernova et al. (1993) noted that although a large part of the differences between low-$P_{max}$ and high-$P_{max}$ comets can be attributed to the contamination by molecular emissions, real differences in properties of dust particles can affect the polarization too.

As it was shown in Kolokolova et al. (2007), the dust in low-$P_{max}$ and high-$P_{max}$ comets is really different, provoking different effect of the gas emission. In high-$P_{max}$ comets, dust almost homogeneously distributed in the coma, whereas in the low-$P_{max}$ comets it is concentrated near the nucleus, and, thus, at the measurements with large apertures, facilitates higher contribution of the gas emissions as the main contribution into the observed polarization comes from the large distances from the nucleus. Such a difference in the distribution of the dust within the coma results from different size or porosity of particles that makes particles in low-$P_{max}$ comets more massive and, thus, concentrating near the nucleus. This is consistent with the IR observations of comets, which show that the high-polarization comets have strong IR excesses and silicate features, while the low-polarization comets show low or absent IR excesses, and weak or absent silicate emissions (Chernova et al., 1993; Levasseur-Regourd et al., 1996; Hanner, 1980; Kolokolova et al., 2007).

Another point of view is that the observed difference in the maximum polarization of comets is due to different composition of their dust. Zubko et al. (2016) totally rejecting the gas contamination effect, believe that the maximum polarization range from 7% up to >30% can be reproduced by the variations in the relative abundance of weakly absorbing particles (e.g., silicates) and highly absorbing particles (e.g., carbonaceous materials). However, their model does not explain the absence of comets with the maximum polarization noticeably below ~8%, whereas this lowest limit of the observed polarization is consistent with the maximum degree of polarization of molecular emissions equal to 7.7%. Also, hypothesis by Zubko et al. cannot provide a reasonable explanation of observed changes in polarization with the distance from the nucleus.

Levasseur-Regourd et al. (1996) noted that comets exhibiting a rather large polarization at the maximum seem to be so-called active comets. It is possible that the increase of the whole coma polarization in the high-$P_{max}$ comets is caused by a significant contribution from polarization of the dust jets. However, as was shown by Jockers et al. (1997), the polarization of dust jets was only slightly higher (1–2%) than that in the surrounding areas of the coma of comet C/1995 O1 (Hale-Bopp). The permanently acting at least one jet (fan) in comet Encke allows us to attribute this comet to the active class, but its polarization in the whole coma is small. Thus, activity of comets does not directly determine their polarization. Consequently, the activity of the comet does not directly define the polarization class of comets. It is possible that the polarization properties of comets are related to a difference in the properties of the dust lifted from the nucleus surface and the dust ejected from the active regions of the nucleus, however, confirmation of this requires further examination.

Summing up, we can say that the observed differences in the polarization of different comets, which are active close to the Sun ($r < 2$ au), without the gas contamination correction (and maybe in some cases without the nucleus contamination correction) cannot unambiguously reveal differences in the properties of their dust particles.

*Distribution of polarization and color over the coma*

In addition to the differences in the cometary polarization for the whole coma (see Fig. 9), variations of polarization in different regions of the coma (Fig. 8) are of a special interest. The behavior of polarization and color with the distance from the nucleus of comet Encke is similar to their behavior observed in comet 67P/C-G (Rosenbush et al., 2017), although the quantitative changes in polarization and color with the distance from the nucleus of both comets are different. This may be due to different properties of the nuclei of two comets and partially the conditions of their observations: phase angle of comet Encke was $\alpha = 46.8°$, and comet 67P/C-G $\alpha = 33.2°$.

As we can see in Fig. 4, at optocentric distances up to ∼3.000 km, the $F_{dust}/F_{gas}$ ratios (that is $F_{cont}/F_{em}$) in the r-sdds and SED500 filters are ∼1 and ∼0.25, respectively. Therefore, a sharp decrease in the observed polarization at these distances (see Figs. 3 and 5) is mainly caused by a change in the $F_{dust}/F_{gas}$ ratio and partly due to the influence of the nucleus. At larger distances from the nucleus ($\rho > 3.000$ km), the observed polarization behavior is more complex. The wave-like polarization changes are superimposed on a gradual increase with distance from the nucleus. They are almost identical for both filters, i.e. they are not caused by the observation errors. At the same distances, the $F_{dust}/F_{gas}$ ratio also increases. This is clearly visible for the r-sdss filter (Fig. 4) but is weakly manifested in the SED500 filter because the $C_2(0,0)$ emission is stronger and extends at a larger distance from the nucleus (see Fig. 4). The increase in the $F_{dust}/F_{gas}$ ratio can be explained by the disintegration of particles into smaller ones, resulting in the increase in the scattering cross section, that in turn explains the increase in the degree of polarization at distances large than 3000 km.

The polarization of the dust coma in the near-nucleus region free from the gas and nucleus contributions decreases from ∼13% up to ∼4% (see Fig. 8). At these distances, there is a sharp decrease in the color BC–RC' from ∼1.4$^m$ to ∼0.4$^m$, which is practically unaffected by molecular emissions, since the narrowband filters transmit the continuum, and the $NH_2$ emission falling into the RC filter was subtracted.

Similar variations of polarization and color were observed in comet 67P/C-G (Rosenbush et al., 2017): the dust color (g–r)$_{sdss}$ gradually decreased with the distance from the nucleus, from 0.8$^m$ to 0.5$^m$, at the distance about 40,000 km. However, the color of Encke's dust (in terms of the reddening per 1000 Å, see Section 9) in the near-nucleus area of the coma is noticeably redder (reddening is ∼15%/1000 Å) than that (8.2%/1000 Å) for comet 67P/C-G, the color change of the dust ("blueing") and the increase of polarization degree towards the outer part of the coma in comet Encke occur faster than those in comet 67P/C-G. Nevertheless, the general trends in the radial variations of polarization and color in comets Encke and 67P/C-G are similar and suggest changing particle properties on a time-of-flight timescale. The scenario considered in detail for comet 67P/C-G in Rosenbush et al. (2017), can also be applied to comet Encke. It is the following. At the nucleocentric distances below ∼3.000 km, a decrease of polarization degree with decreasing particle size is expected when the size of particles decreases from hundreds of microns to some microns. Then the polarization start increasing that can be attributed to the further decrease in particle size. Such a behavior of polarization is consistent with the laboratory measurements of light scattering by aggregated particles (Hadamcik and Levasseur-Regourd, 2009). These experiments showed that a decrease in size for particles in the range of sizes between 4 and 400 led to decrease in polarization, which changed to an increase in polarization as the particles become even smaller approaching the Rayleigh regime. This trend did not depend on the particle composition and was typical for silicates, graphite, and tholin particles. Decrease in particle size is also consistent with the change of the color index over the coma: it progressively decreased, although did not become negative at periphery of the coma indicating absence of a significant amount of very small, Rayleigh particles. These conclusions are also confirmed by a simple Mie modeling in Rosenbush et al. (2017) and by more sophisticated Sh-matrix modeling of Gaussian particles in this paper.

**Conclusion**

In this paper we presented the results of the polarimetric observations of comet 2P/Encke performed at the 6-m BTA telescope of the Special Astrophysical Observatory (Russia) on January 23, 2017. These results are compared with the color index and the dust/gas flux ratio and the contribution of light scattered by the nucleus into the total flux from the coma derived in Paper I and computer simulation of change in color and polarization. The obtained results allow us to draw the following main conclusions:

1. Dust in Encke is mainly concentrated in the near-nucleus region of the coma. The maximum $F_{cont}/F_{em}$ ratios are 1.48 and 2.91 in the SED500 and the r-sdss filters but dropped sharply to ∼0.2 and ∼1 at the distance ∼2.500 km, respectively. Nevertheless, dust expands to large distances (>12,000 km) from the nucleus that is manifested by an increase of the dust/gas ratio up to ∼0.3 (SED500) and ∼1.3 (r-sdds) and by a high degree of polarization equal to about 11–12% in both filters.

2. The maximum degree of polarization, ∼8% in the r-sdss filter, was detected in the dust shell (or cloud) which is located at ∼1.000 km towards the Sun.

3. The observed polarization in the near-nucleus area of the coma was ∼6% and then dropped sharply to ∼4% at the distance ∼2.000 km in the SED500 filter; in the r-sdss filter it was ∼8% near the nucleus and dropped also to ∼4% at ∼2.000 km. At larger distances from optocenter, the polarization over the coma gradually increased with wave-like fluctuations, reaching ∼6% (SED500 filter) and ∼8% (r-sdss filter) at the distances about 13,000 km. After correction for the gas contamination, using the dust/gas ratio from the spectroscopy, the gas-free polarization degree of the dust increased by ∼2% in SED500 and ∼5% in r-sdss band in the near-nucleus coma, and at the distance about 12,000 km the increase was by ∼6% and ∼3%, respectively.

4. The effect of the light scattered by the nucleus on the polarization of the dust coma was first studied in comets. This effect for comet Encke was equal about 0.7% and appears directly near the cometary nucleus, at distances ≤ 1.000 km.

5. The maximum polarization in the near-nuclear region of the coma corrected for gas contamination is in a good agreement with the synthetic phase-angle dependence of polarization for the whole coma of the high-$P$-max comets. This proves that the gas contamination really affects the polarization values, and this may be one of the reasons of the diversity in the polarization of comets at large phase angles.

6. The significant variations of polarization and color of dust in comet Encke are very similar to those in comet 67P/C-G. This suggests some evolution of particle properties as they move out of the nucleus. Our modeling, using Sh-matrix method for Gaussian particles, demonstrates that the observed radial trends in color and polarization can be explained by decreasing size of the dust particles with the distance from the nucleus.

Thus, the degree of polarization and its change in the coma of comet Encke are influenced by the concentration of dust in the near-nucleus region, the depolarization effect of gas emissions that depends on the dust/gas ratio, the polarization of light scattered by the nucleus, although in lesser degree, and the change in physical properties of the dust particles at their journey through the coma.

**Acknowledgments**

We thank the anonymous referees for very careful reading of the manuscript and their helpful comments. The observations at the 6-m

BTA telescope were performed with the financial support of the Ministry of Education and Science of the Russian Federation (agreement No. 14.619.21.0004, project ID RFMEFI61914X0004). The authors express appreciation to the Large Telescope Program Committee of the RAS for the possibility of implementing the program of observations of comet 2P/Encke at the BTA. The researches by VR, OI, and VK are supported, in part, by the project 16BF023-02 of the Taras Shevchenko National University of Kyiv. Researches by NK and DP were funded by Russian Foundation for Basic Research and the government of the Crimean Republic, grant № 18-42-910019\18. OI thanks the SASPRO Programme and the Slovak Academy of Sciences (grant Vega 2/0023/18). The researches by VR, OI, and NK were supported, in part, by the Ukrainian-Slovak joint research project for the period 2017–2019.